\documentclass[acmsmall, screen]{acmart}

\AtBeginDocument{%
  }

\setcopyright{acmcopyright}
\copyrightyear{2022}
\acmYear{2022}
\acmJournal{JACM}
\acmArticle{1}
\usepackage[ruled,vlined,linesnumbered]{algorithm2e}
\newcommand{\eqbf}[1]{\boldsymbol{#1}}
\usepackage{algorithmic}
\usepackage{multirow}
\begin{document}

\title{Collaborative Group-Aware Hashing for Fast Recommender Systems}

\author{Yan Zhang}
\email{yan.zhang@cdu.edu.au}
\affiliation{%
	\institution{Faculty of Science and Technology, Charles Darwin University}
%	\streetaddress{xxx}
	\city{Darwin}
	\state{NT}
	\country{Australia}
	\postcode{0800}
}
% \affiliation{%
% 	\institution{ University of Technology Sydney}
% 	%	\streetaddress{xxx}
% 	\city{Sydney}
% 	\state{NSW}
% 	\country{Australia}
% 	\postcode{2007}
% }

\author{Li Deng}
\email{oliverdengli@gmail.com}
\affiliation{%
	\institution{School of Computer Science and Engineering, University of Electronic Science and Technology of China}
%	\streetaddress{xxx}
	\city{Chengdu}
	\state{Sichuan}
	\country{China}
	\postcode{611731}
}

\author{Lixin Duan}
\email{lxduan@gmail.com}
\affiliation{%
	\institution{School of Computer Science and Engineering, University of Electronic Science and Technology of China}
%	\streetaddress{xxx}
	\city{Chengdu}
	\state{Sichuan}
	\country{China}
	\postcode{611731}
}

\author{Ivor W. Tsang}
\email{Ivor_Tsang@cfar.a-star.edu.sg}
\affiliation{%
  \institution{CFAR and IHPC, Agency for Science, Technology and Research, Singapore College of
Computing and Data Science, Nanyang Technological University, Singapore}
%  \city{xxx}
  \country{Singapore}
}

% \author{Wen Li}
% \email{liwenbnu@gmail.com}
% \affiliation{%
% 	\institution{School of Computer Science and Engineering, University of Electronic Science and Technology of China}
% %	\streetaddress{xxx}
% 	\city{Chengdu}
% 	\state{Sichuan}
% 	\country{China}
% 	\postcode{611731}
% }
\author{Guowu Yang}
\email{guowu@uestc.edu.cn}
\affiliation{%
	\institution{School of Computer Science and Engineering, University of Electronic Science and Technology of China}
	%	\streetaddress{xxx}
	\city{Chengdu}
	\state{Sichuan}
	\country{China}
	\postcode{611731}
}

\renewcommand{\shortauthors}{Yan et al.}

\begin{abstract}
The fast online recommendation is critical for applications with large-scale databases; meanwhile, it is challenging to provide accurate recommendations in sparse scenarios. Hash technique has shown its superiority for speeding up the online recommendation by bit operations on Hamming distance computations. However, existing hashing-based recommendations suffer from low accuracy, especially with sparse settings, due to the limited representation capability of each bit and neglected inherent relations among users and items. To this end, this paper lodges a Collaborative Group-Aware Hashing (CGAH) method for both collaborative filtering (namely CGAH-CF) and content-aware recommendations (namely CGAH) by integrating the inherent group information to alleviate the sparse issue. Firstly, we extract inherent group affinities of users and items by classifying their latent vectors into different groups. Then, the preference is formulated as the inner product of the group affinity and the similarity of hash codes. By learning hash codes with the inherent group information, CGAH obtains more effective hash codes than other discrete methods with sparse interactive data. Extensive experiments on three public datasets show the superior performance of our proposed CGAH and CGAH-CF over the state-of-the-art discrete collaborative filtering methods and discrete content-aware recommendations under different sparse settings.

\end{abstract}

\begin{CCSXML}
	<ccs2012>
	<concept>
	<concept_id>10002951.10003317.10003347.10003350</concept_id>
	<concept_desc>Information systems~Recommender systems</concept_desc>
	<concept_significance>500</concept_significance>
	</concept>
	<concept>
	<concept_id>10002951.10003317.10003347.10003352</concept_id>
	<concept_desc>Information systems~Information extraction</concept_desc>
	<concept_significance>500</concept_significance>
	</concept>
	<concept>
	<concept_id>10002951.10003317.10003331.10003271</concept_id>
	<concept_desc>Information systems~Personalization</concept_desc>
	<concept_significance>500</concept_significance>
	</concept>
	<concept>
	<concept_id>10002951.10003317.10003338.10003346</concept_id>
	<concept_desc>Information systems~Top-k retrieval in databases</concept_desc>
	<concept_significance>500</concept_significance>
	</concept>
	<concept>
	<concept_id>10002951.10003317.10003338.10003342</concept_id>
	<concept_desc>Information systems~Similarity measures</concept_desc>
	<concept_significance>500</concept_significance>
	</concept>
	</ccs2012>
\end{CCSXML}

\ccsdesc[500]{Information systems~Recommender systems}
\ccsdesc[500]{Information systems~Information extraction}
\ccsdesc[500]{Information systems~Personalization}
\ccsdesc[500]{Information systems~Top-k retrieval in databases}
\ccsdesc[500]{Information systems~Similarity measures}

\keywords{Recommender system, hash code, group affinity, content-aware recommendation}

\maketitle

\section{Introduction} \label{sec1}
\vspace{0.5em}
Recommender systems have been extensively applied in the growing number of e-commerce websites for helping their customers find desirable products. Recommender systems recommend a particular user with a small set items from the underlying pool of items that the user may be interested in. Most of the existing recommendation models were based on users' previous behavior data such as ratings, purchasing records, click actions, and watching records, like/dislike records, etc. One of the most successful Collaborative Filtering (CF) is Matrix Factorization (MF). MF maps users and items into $r$-dimensional real latent space, and then provides recommendations by ranking inner products between a specific user and all items. MF-based methods have achieved excellent recommendation accuracy in both academia and industrial. However, it's challenging to provide a fast response for online recommending with large number of items in the real latent space \cite{zhang2014preference,zhou2012learning}.

%The emerge of hashing-based recommendations makes online recommendation efficient on both response time and storage costs. Several scholars put forward hashing-based recommendation frameworks~\cite{zhou2012learning,zhang2016discrete,zhang2017discrete} twhat significantly accelerate the online recommendation. To be specific, the hashing-based recommendation framework firstly encodes users and items into binary codes in a common $r$-dimensional Hamming space. Then preference scores denoted by Hamming distances can be efficiently computed by bit operations. One can even use a fast and accurate indexing method to find approximate top-\emph{k} preferred items with sublinear or logarithmic time complexity~\cite{wang2012semi}. So the time cost with hashing-based frameworks is significantly reduced compared with continuous frameworks. In addition, each dimension of hash codes can be stored with only one bit instead of 32/64 bits that are used for storing one dimension of continuous vectors, which greatly reduces storage cost as well. 

Due to observed ratings in the real world are generally sparse, which leads to poor performance with CF based recommender systems. Existing work for solving sparse and cold start issues mainly includes three folds: (1) MF variants by incorporating latent vectors extracted from side information \cite{koren2008factorization,mairal2010online, lian2016regularized,wang2011collaborative,wang2015collaborative,xu2018graphcar}, which improves the recommendation accuracy with the aid of side information, such as items' figures, users' profile, and the social network of users, et al. Specifically, content-aware recommender systems learn representations of users and items from their content data, and then these representations were unified into a CF-based framework, and thus it can be applied to cope with data sparsity problem. (2) Deep models of formulating projections from side information to interactive data \cite{he2017neural,zhang2019deep,he2017neural2,xue2017deep}. (3) Cross-modal models of transferring knowledge from a domain with rich information to another domain with very sparse historical data \cite{zhao2017unified,ding2019learning,nguyen2017personalized,he2018robust}. 

However, the content-aware recommendations learn continuous representations which hinders to provide a fast response for online recommending with large number of items in the real latent space \cite{zhang2014preference,zhou2012learning}, because similarity search in continuous space is time consuming when it comes across millions or billions of items. The emerge of hashing-based recommendations makes online recommendation efficient on both response time and storage costs. Several scholars put forward hashing-based recommendation frameworks~\cite{zhou2012learning,zhang2016discrete,zhang2017discrete,ali2025communication,hu2022artificial,wu2023graph} that significantly accelerate the online recommendation. To be specific, the hashing-based recommendation framework firstly encodes users and items into binary codes in a common $r$-dimensional Hamming space. Then preference scores denoted by Hamming distances can be efficiently computed by bit operations. One can even use a fast and accurate indexing method to find approximate top-\emph{k} preferred items with sublinear or logarithmic time complexity~\cite{wang2012semi}. So the time cost with hashing-based frameworks is significantly reduced compared with continuous frameworks. In addition, each dimension of hash codes can be stored with only one bit instead of 32/64 bits that are used for storing one dimension of continuous vectors, which greatly reduces storage cost as well. 
%So some hybrid hashing-based recommendations are proposed to improve both the accuracy and efficiency for the online recommendation under sparse settings. 

%However, it's challenging to provide a fast response for online recommending with large number of items in the real latent space \cite{zhang2014preference,zhou2012learning}.

\begin{figure}[!htbp]
	%	\centering
	\includegraphics[width=1\linewidth]{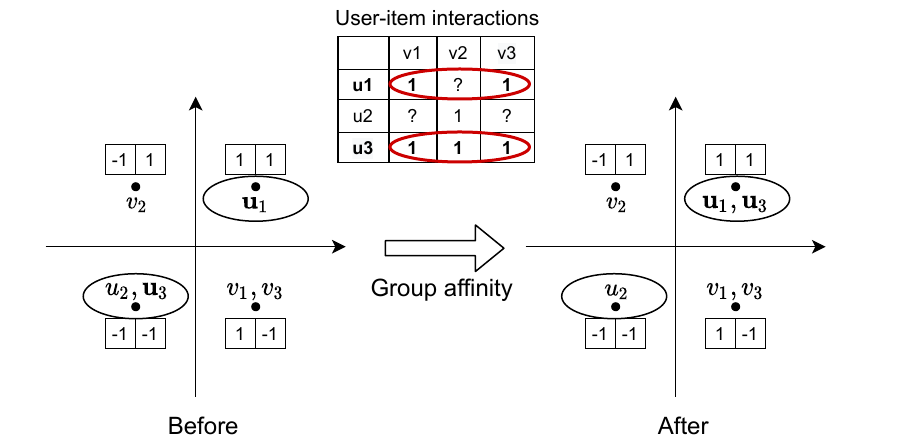}
	%		\subfigure{\includegraphics[width=1\linewidth]{mzf}}
	\caption{The motivation of our method. From the top ‘User-item interactions’, we find user $u_1$ has similar behaviors with user $u_3$, so we hope to learn similar hash codes of users $u_1$ and $u_3$. Before considering inherent group information, we obtain different hash codes with DCF \cite{zhang2016discrete}. After considering group affinity, we can obtain the same or more similar hash codes for $u_1$ and $u_3$ by using the proposed CGAH}\label{fig}
	%\vspace{-1em}
\end{figure}

To improve both the accuracy and efficiency with sparse settings, many scholars proposed hashing-based recommendation schemes with content information \cite{lian2017discrete,lian2020lightrec,hansen2020content,liu2019discrete,lu2019learning}. Most existing hashing frameworks learned hash codes from users' observed historical behavior data and side information, which neglects the inherent relation among users or items, which leads to unsatisfactory recommendations. For example, NeuHash-CF \cite{hansen2020content} was modeled as an autoencoder architecture consisting of two joint hashing components for generating user and item hash codes. NeuHash-CF works in sparse and cold-start settings, but hash codes learned from two joint hashing components with autoencoders cannot capture the inherent group relation amongst users and items, which leads to hash codes of similar items/users are different from each other as shown in Figure. \ref{fig}. We hope to learn hash codes that are consistent with the group affinities in the original latent space of users and items. The idea of considering group affinity to formulate our model comes from the memory-based collaborative filtering in sparse settings, where a user's preference can be predicted by their neighbors. 

Actually, in the real world applications, people's preferences over items have evident group features. The inherent group relationship is naturally exists in the real world. For example, users with similar characteristics like similar movies or books. Similarly, items with similar abstract information or category information are appealing to a group of users. Especially, the rise of social media has made group activities very common in social applications. For example, we often join a group to buy a product on Pinduoduo \footnote{https://en.pinduoduo.com/}, and join a group to travel together on Tuniu \footnote{https://www.tuniu.com/}. This motivates us to study behaviors of groups, which in turn benefit us to develop a recommendation for coping with the data sparsity and cold-start issues. 

To this end, we propose a collaborative group-aware hashing method to improve both the accuracy in sparse settings and the efficiency of online recommendation for large-scale datasets. Motivated by the memory-based collaborative filtering methods in sparse settings, where a user's preference can be predicted by their neighbors. Besides, MF maps users and items into a shared $r$-dimensional latent space and predicts preference scores with their inner products, but it neglects inherent relations within users and items. The inherent relationship is naturally exists in the real world, so that users with similar characteristic like similar movies or books, similarly, items with similar abstract information is appealing to one group of users. In this paper, we seek an efficient way to improve the accuracy with inherent group information in sparse settings. It is expected to assist recommendations for users/items with just a few ratings. Our contributions are summarized as follows:
%Meanwhile, to speed up the online recommendation, we employ binary codes to represent users and items. 
%factorizes a rating matrix into a user matrix and an item matrix, but it only considers the complex relationship to predict the unobserved ratings, and does not attain more abstract representations (such as the type of movies: romance, action, drama, et al ). 
\begin{itemize}
	\item We propose an efficient collaborative group-aware hashing (CGAH) method for recommendations, which encodes users and items with hash codes by incorporating inherent group affinities neglected by the state-of-the-arts hashing frameworks. To the best of our knowledge, this is the first hashing-based scheme which explicitly learns hash codes with the inherent group information.
	\item We build a novel hashing model for recommendations by formulating the preference model with the group affinity and the similarity of hash codes in Hamming space. Besides, the learned hash codes are in line with group affinities, which enhances the representation capability of hash codes and favors to providing more accurate recommendations under sparse settings.
	\item We develop an efficient discrete optimization method to solve the proposed objective.
	\item We design extensive experiments on three public datasets to show the superiority of CGAH and CGAH-CF on both accuracy and efficiency over the state-of-the-art discrete CF-based methods and the discrete content-aware recommendations under different sparse environments.
\end{itemize}

\section{Related work} \label{sec2}
\vspace{0.5em}
%A pioneer work~\cite{das2007google} was proposed to exploit Locality-Sensitive Hashing~\cite{datar2004locality} to generate hash codes for Google News readers based on their click history. On this basis, A. Karatzoglou et al.~\cite{karatzoglou2010collaborative} randomly projected real latent representations learned from regularized matrix factorization into hash codes. Similar to this, K. Zhou et al.\cite{zhou2012learning} followed the idea of Iterative Quantization~\cite{gong2013iterative} to generate binary codes from rotated real latent representations. To derive compact binary codes, the uncorrelated constraints were imposed on the real latent representations in regularized matrix factorization. However, according to the analysis in~\cite{zhang2014preference}, hashing essentially only preserves similarity rather than preference based on inner product, since the magnitudes of representations for users and items are discarded in the quantization stage. Thus, a constant feature norm (CFN) constraint was imposed when learning the real latent representations, and then the magnitudes and similarity are respectively quantized in~\cite{zhang2017dot}. But the two-stage approach still suffered large information loss in the quantization procedure. 

\section{Related Work}

\subsection{Content-aware Recommendations}
\vspace{0.5em}
Content-aware recommender systems have been extensively studied to cope with data sparsity and cold-start challenges by integrating side information, such as users' demographic data, items' descriptions and tags, users' reviews, users'/items' images, relations of users/items, etc. Such as the state-of-the-art collaborative deep learning (CDL)~\cite{wang2015collaborative} was proposed as a probabilistic model by jointly learning a probabilistic stacked denoising auto-encoder (SDAE)~\cite{vincent2010stacked} and CF. CDL exploits the interaction and content data to alleviate cold start and data sparsity problems. However, unlike CTR, CDL takes advantage of deep learning framework to learn effective real latent representations. Thus, it can also be applied in the cold-start and sparse settings. CDL is also a tightly coupled method for recommender systems by developing a hierarchical Bayesian model.

Another classic content-aware method is deep cooperative neural networks (CoNN)\cite{zheng2017joint}, which consists of two parallel neural networks: one learns user behaviors exploiting users' reviews, and the other one learns  item properties from the items' reviews. A shared layer is introduced on the top to couple these two networks together. Then, dual attention mutual learning (DAML)\cite{liu2019daml} integrates ratings and reviews into a joint neural network with a local and mutual attention mechanism to strengthen the interpretability. In addition, higher-order nonlinear interaction of features are extracted by the neural factorization machines to predict ratings. 

\subsection{Group-aware Recommendations}
\vspace{0.5em}
As group activity is becoming popular in our life. Recommendations for a group of users becomes significant in current social medias. In fact, group-aware recommendations is also meaningful in traditional information system, because it can alleviate cold-start and sparse issues by group information. Many researchers focus on group recommendations, such as Agalya et al. \cite{agalya2016group} proposed a top-k 
recommendation scheme by the interrelationship of each group, which makes the recommendation efficient and accurate, that can be applied in Movies, Product, and Travel Recommendations. Based on NCF, Cao et al. \cite{cao2018attentive} proposed an attention mechanism to learn groups' representations. They combine interactions of individual users on items of groups and user-item interactions into their model to improve the performance. With the rise of sequential recommendations, Wang et al. \cite{wang2020group} combined sequential recommendation and group recommendation to learn dynamic group representations by proposing a novel sequential group recommendation method, which is effective to obtain good performance. Specifically, they devised a Long-term Group-aware and
Short-term Graph-aware Representation Learning approach and for sequential-based group recommendations, a long-term and a short-term group-aware graph to capture user-item interactive information. As current group recommendation frameworks fail to extract complex relations among users and items, Liang et al. \cite{liang2021hierarchical} proposed a hierarchical fuzzy graph attention network to
improve fuzzy profiling for both groups and items. 

\subsection{Hashing for Collaborative Filtering}
\vspace{0.5em}
hashing-based recommender systems contains two types: two-stage hashing and learning based hashing. One representative work of two-stage hashing methods is Binary Codes for Collaborative Filtering (BCCF) \cite{zhou2012learning}, which learn the real latent vectors of users and items, and then quantize them into binary codes. BCCF improves the efficiency of the online recommendation with binary codes, but it suffers poor accuracy performance due to the large information loss in the quantization stage. 

Different from two-stage hashing frameworks, learning based hashing methods learn hash codes by directly optimizing discrete problems. Thus more information is carried by hash codes than two-stage frameworks. Zhang et al. proposed discrete collaborative filtering (DCF)~\cite{zhang2016discrete}. By adding balance and uncorrelated constraints on hash codes, DCF obtained efficient binary codes. DCF was evaluated using a similar way of the conventional CF~\cite{rendle2009bpr}. Then, a ranking-based hashing framework was proposed to improve the recommendation performance~\cite{zhang2017discrete} by adding the same constraints with DCF, and it also obtained short and informative hash codes. To improve the accuracy of discrete recommendation, Wang et al. proposed the adversarial binary collaborative filtering (ABinCF) \cite{wang2019adversarial} formulated as a min-max optimization problem. Liu et al. proposed the compositional coding for collaborative Filtering (CCCF) \cite{liu2019compositional} which innovatively represents each user/item with a set of binary vectors, which are associated with a sparse real-value weight vector. Each value of the weight vector encodes the importance of the corresponding binary vector to the user/item. The continuous weight vectors greatly enhances the representation capability of binary codes, and its sparsity guarantees the processing speed.

\subsection{Hashing for Content-aware Recommendations}
\vspace{0.5em}
%To alleviate sparse and cold start issues, Lian et al. \cite{lian2020lightrec} proposed LightRec in which each item was represented as additive composition of B codewords, which are optimally selected from each of the codebooks. LightRec and CCCF exploits similar compositional coding methods to present users and items, but LightRec utilizes a shared B codewords for improving the efficiency on both time and storage.  Another type of hashing-based recommendations are developed with deep models, such as NeuHash-CF \cite{hansen2020content} was formulated as an autoencoder architecture consisting of two joint hashing components for generating user and item hash codes. Liu et al. proposed a discrete social recommendation (DSR) \cite{liu2019discrete} method by learning binary codes in a unified framework of users and items considering social information. To take dual advantages of both social relations and discrete technique for fast recommendation, Discrete Trust-aware Matrix Factorization (DTMF) \cite{guo2019discrete} was proposed by learning latent features of truster and trustees based on trust-aware matrix without sacrificing too much information. To solve the cold-start and sparsity issue, Multi-Feature Discrete Collaborative Filtering (MFDCF) \cite{xu2020multi} was proposed by projecting multiple content features of users into binary hash codes via fully exploiting their complementarities. 

By combining with the content information, Zhang et al. presented the Discrete Deep Learning (DDL) \cite{zhang2018discrete1} for solving cold start and spare problems. Besides, Liu et al. proposed a discrete social recommendation (DSR) \cite{liu2019discrete} method by learning binary codes in a unified framework of users and items by considering social information. To take dual advantages of both social relations and discrete technique for fast recommendation, Discrete Trust-aware Matrix Factorization (DTMF) \cite{guo2019discrete} was proposed by learning latent features of truster and trustees based on trust-aware matrix without sacrificing too much information. Then, Lian et al. \cite{lian2020lightrec} proposed LightRec, where each item was represented as additive composition of B codewords, which are optimally selected from each of the codebooks. LightRec and CCCF exploits similar compositional coding methods to present users and items, but LightRec utilizes a shared B codewords for improving the efficiency on both time and storage.  Another type of hashing-based recommendations are developed with deep models, such as NeuHash-CF \cite{hansen2020content} was formulated as an autoencoder architecture consisting of two joint hashing components for generating user and item hash codes. To solve the cold-start and sparsity issue, Multi-Feature Discrete Collaborative Filtering (MFDCF) \cite{xu2020multi} was proposed by projecting multiple content features of users into binary hash codes via fully exploiting their complementarities. 
%%\vspace{-6pt}
\section{Collaborative Group-Aware Hashing (CGAH)}
\vspace{0.5em}
In this section, we first formulate the problem/task of this paper in Section \ref{sec30}. Then we present the preliminary study related to our proposed model in Section \ref{sec31}. Next, we introduce the proposed Collaborative Group-Aware Hashing (CGAH) for collaborative filtering and content-aware recommendations separately in Section \ref{sec32} and Section \ref{sec33}. Finally, we develop an alternating optimization strategy to solve the mix-integer programming problems of CGAH in Section \ref{sec34}.
%%\vspace{-6pt}

\begin{table}%[!t]
	%\small
	\caption{Notations}
	\centering
	\begin{tabular}{ll}
		\toprule[1pt]
		Symbol  & Description \\
		\midrule[1pt]
		$I$ & users' index set\\
		$J$ & items' index set\\
		$\eqbf{R}$  & binary matrix of all users in $I$ \\
		$\eqbf{C}_{i}$  & the content data of user $i$  \\
		$\eqbf{C}_{j}$  & the content data of item $j$  \\
		$\eqbf{H}$  & real latent vectors of all users in $I$ \\
		$\eqbf{G}$  & real latent vectors of all items in $J$ \\
		$\eqbf{h}_{i}$  & real latent vector of user $i$ \\
		$\eqbf{g}_{j}$  & real latent vectors of item $j$ \\
		$\eqbf{B}$  & hash codes of all users in $I$ \\
		$\eqbf{D}$  & hash codes of all items in $J$ \\
		$\eqbf{b}_{i}$  & hash code of user $i$ \\
		$\eqbf{d}_{j}$  & hash code of item $j$ \\
		$\mathcal{K}$ & centroids ($k$-dimensional vectors) in the latent space of $\eqbf{H}$ and $\eqbf{G}$ \\
		$\eqbf{P}$  & group indicators of  $\eqbf{H}$ \\
        $\eqbf{Q}$  & group indicators of $\eqbf{G}$ \\
        $\eqbf{p}_{ik}$  & the distance of $\eqbf{h}_{i}$ and $\mathcal{K}_{k}$ \\
        $\eqbf{p}_{ik}$  & the distance of $\eqbf{g}_{j}$ and $\mathcal{K}_{k}$ \\
        $s_{ij}^{k}$ & the group affinity of user $i$ and item $j$ \\
        $\eqbf{\xi}_{i}$ & the embedding vector from the content $\eqbf{C}_{i}$ of user $i$ \\
		$\eqbf{\zeta}_{j}$ &the embedding vector from the content $\eqbf{C}_{j}$of item $j$  \\
		$\Theta_{I}$ & the parameters of encoding network of DAE for users \\%that contains
		$\Theta_{J}$ & the parameters of encoding network of DAE for items \\
		$\eqbf{X}$  & the delegated real matrix of $\eqbf{B}$ \\
		$\eqbf{Y}$  & the delegated real matrix of $\eqbf{D}$ \\
		\bottomrule[1pt]
	\end{tabular}
	\label{tab:1}
\end{table}

\subsection{Problem Formulation} \label{sec30}
\vspace{0.5em}
Let the index sets of users and items be denoted as $I=\{1,\cdots ,n\} $ and $J=\{1\cdots ,m\}$, respectively. $r_{ij} \in \eqbf{R}$ denotes the rating of user $i$ rated to item $j $. $\eqbf{C}_{i}$ and $\eqbf{C}_{j}$ denotes content data of user $i$ and item $j$, respectively. The detailed notations used in this paper are listed in Table \ref{tab:1}

Our tasks includes: 
\begin{itemize}
	\item[(1)] Given ratings $\eqbf{R}$, we firstly capture group affinities of users and items. Our objective is to learn hash codes of users and items by solving a discrete optimization problem, which is formulated with group affinities, ratings, and parameters; and then obtain top-$k$ recommendations based on learned hash codes. We call the method as 'collaborative group-aware hashing for collaborative filtering' (CAGH-CF). 
	\item[(2)] Given ratings $r_{ij}\in \eqbf{R}$ and content $\eqbf{C}_{i}$, $\eqbf{C}_{j}$ for user $i$ and item $j$. Our objective is to capture group affinities of users and items from both ratings and content; and learn hash codes of users and items by solving a discrete optimization problem formated with group affinities and ratings; and finally recommend top-$k$ items for users. We also name the method as 'CGAH for content-aware recommendations' (CGAH).
\end{itemize}

\subsection{Preliminary} \label{sec31}
\vspace{0.5em}
Collaborative filtering (CF) provides recommendations based on user-item interactive data, i.e., ratings used in this paper. Suppose each entry $r_{ij}$ of the rating matrix $\eqbf{R} \in \mathbb{R}^{n\times m}$ denotes the preference of user $i$ to item $j$. The traditional real-valued CF methods encodes users and items with $r-$dimensional real latent vectors $\eqbf{h}_{i}$ and $\eqbf{g}_{j}$ by matrix factorization (MF), and then the unobserved preferences $\hat{r}_{ij}$ are estimated by the inner product of the corresponding real latent vectors $\hat{r}_{ij}=\eqbf{h}_{i}^{T}\eqbf{g}_{j} $.  Thus the MF model is generally formulated as
\begin{equation}\label{mf}
\arg \min_{\eqbf{H}, \eqbf{G}} \mathcal{L}_{\tiny{MF}}(\eqbf{R}, \eqbf{H}^{T}\eqbf{G})+\lambda \Gamma(\eqbf{H}, \eqbf{G}),
\end{equation}
where $\eqbf{H}\in \mathbb{R}^{r\times n}, \eqbf{G}\in \mathbb{R}^{r\times m}$. The first term is the loss function of the true ratings and the predicted ratings. The loss functions of existing MF-based invariants consist of two types: (1) rating-based losses \cite{koren2009matrix,ma2008sorec,ma2008sorec,wang2015collaborative,zhang2014preference} and ranking-based losses \cite{weimer2008cofi,weimer2007maximum,rendle2009bpr,tang2018ranking}. The second term is the regularizer of latent vectors $\eqbf{H}$ and $\eqbf{G}$, which is applied to avoid overfitting. Recommendations based on MF have achieved good performance in both academia and industrial, however, the efficiency of online recommendation has becoming more and more crucial with the increasing numbers of items and users. Fortunately, hashing has been shown as one of the most promising technique for providing efficient online recommendations.

The state-of-the-art discrete CF-based methods encode users and items into $r-$dimensional binary codes $\eqbf{b}_{i}$ and $\eqbf{d}_{j}$, and then formulate preferences $\hat{r}_{ij}$ with similarities defined by the Hamming distances of hash codes $\hat{r}_{ij}=1-\frac{1}{r}{\text{dis}_{\tiny{H}}(\eqbf{b}_{i},\eqbf{d}_{j})}$. Taking DCF as an example, it is formulated as following objective,
\begin{equation}
\arg \min_{\eqbf{B}, \eqbf{D}} \mathcal{L}_{\tiny{DCF}}(\eqbf{R}, \eqbf{B}^{T}\eqbf{D})+\lambda \Gamma(\eqbf{B}, \eqbf{D}),
\end{equation}
similarly, the loss functions contains rating-based and ranking-based loss functions. Due to the limit representation ability of each bit in hash codes, the discrete CF methods often suffer worse performance than continuous CF models, especially under sparse scenarios.
%%\vspace{-6pt}
\subsection{CGAH for Collaborative Filtering} \label{sec32}
\vspace{0.5em}
Sparsity as one of challenges in recommender system, has becoming more and more serious with the increasing users and items. Thus it's significant to design accurate yet efficient recommendations under sparse settings. Previous studies alleviated the sparsity issue with hybrid recommendation techniques which incorporated content information and interactive data, but very few studies focus on providing both efficient and accurate recommendations only dependent on sparse interactive data. In this section, we attempt to improve the efficiency of CF without degrading the prediction performance. Since the fixed distance between each pair of binary codes impedes them from modeling different magnitudes of relationships within users and items, we assign real-valued weights $s_{ij}$ to each pair to explicitly denote the group affinities of users and items. 

\begin{figure}[t]
	\centering
	\includegraphics[width=0.8\linewidth]{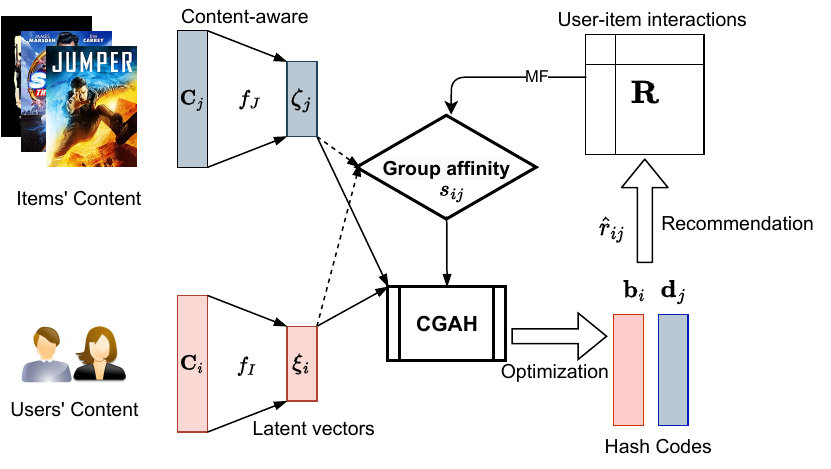}
	%		\subfigure{\includegraphics[width=1\linewidth]{mzf}}
	\caption{The framework of CGAH: the input includes the user content $\eqbf{C}_{i}$, the item content $\eqbf{C}_{j}$, and user-item interactive data, i.e., ratings $\eqbf{R}$; CGAH first extracts latent vectors $\eqbf{h}_{i}$, $\eqbf{g}_j$ from ratings by matrix factorization model, and $\eqbf{\xi}_{i}$, $\eqbf{\zeta}_j$ respectively from users' content and items' content data by encoders; then it concatenates latent vectors from interactive data and content information to attain group infinities $s_{ij}$ in the shared latent space; finally CGAH learns hash codes $\eqbf{b}_i$ and $\eqbf{d}_j$ by solving discrete objectives with group infinities.}\label{fig0}
	%%\vspace{1em}
\end{figure}

As shown in Figure \ref{fig0}, the framework of CGAH consists of two phases: the first phase learns latent vectors $\eqbf{h}_{i}$, $\eqbf{g}_j$ and $\eqbf{\xi}_{i}$, $\eqbf{\zeta}_j$ respectively from interactive data and users'/items' content data, and then concatenates them together to calculate group affinities $s_{ij}$ in the shared latent space; the second phase learns hash codes $\eqbf{b}_i$ and $\eqbf{d}_j$ by solving discrete objectives of CGAH-CF and CGAH for content-aware recommendations with group infinities. We formulate CGAH with two objectives respectively for collaborative filtering and content-aware recommendations.

In this section, we will introduce the model of CGAH for collaborative filtering. Motivated by the memory based CF for the recommendation in sparse settings, where a user's preference can be predicted by their neighbors. Besides, the MF factorizes the rating matrix $\eqbf{R}$ into a user matrix $\eqbf{H}$ and an item matrix $\eqbf{G}$, but it only considers users‘ and items' own affinities in the latent space to predict unobserved ratings, and does not consider the inherent group affinities existed within users and items, such as types of characteristics for users: generosity, integrity, loyalty, devotion, loving, kindness, et al., and abstract types of movies: romance, action, drama, et al. In this paper, we seek an efficient way to improve the recommendation accuracy under sparse settings with the inherent group information. In MF, we suppose users' latent preferences over items' abstract types are classified into $\kappa$ groups, then we denote the real latent vectors of users and items $\eqbf{H}$ and $\eqbf{G}$ as multiplications of group indicator matrices $\eqbf{P} \in \mathbb{R}^{\kappa\times n}, \eqbf{Q} \in \mathbb{R}^{\kappa\times m}$ respectively:
\begin{equation}
\eqbf{H}=\mathcal{K}^{T}\eqbf{P},  \eqbf{G}=\mathcal{K}^{T}\eqbf{Q},
\end{equation}
where $\mathcal{K} \in \mathbb{R}^{\kappa\times r}$ is a shared codebook in the latent space, which is stacked by $\kappa$ vectors in $\eqbf{H}$ and $\eqbf{G}$. To enhance the representation ability of group indicators $\eqbf{P}$ and $\eqbf{Q}$, we impose the uncorrelated constraints on vectors in $\mathcal{K}$ and derive the following objective,
\begin{equation}\label{group}
\begin{aligned}
\arg\min_{\mathcal{K},\eqbf{P}, \eqbf{Q}}&||\eqbf{H}-\mathcal{K}^{T}\eqbf{P}||_{F}^{2}+||\eqbf{G}-\mathcal{K}^{T}\eqbf{Q}||_{F}^{2}, \\
s.t. &\mathcal{K}\mathcal{K}^{T}=r\eqbf{I}_{\kappa}. 
\end{aligned} 
\end{equation}
The problem (\ref{group}) is intractable due to the uncorrelated constraint, thus we seek a way to find an approximate solution. Since group indicators $\eqbf{P}$ and $\eqbf{Q}$ can be considered as distances away from vectors in $\mathcal{K}$. By associating it with K-means clustering, these $\kappa-$dimensional vectors in the shared codebook $\mathcal{K}$ can be regarded as centroids in the latent space of $\eqbf{H}$ and $\eqbf{G}$, and each latent vector can be represented by these centroids. To simplify the calculation of group affinities, we obtain centroids in $\mathcal{K}$ by K-means clustering of vectors in the latent space of $\eqbf{H}$ and $\eqbf{G}$, and then we compute distances $\eqbf{P}$ and $\eqbf{Q}$ away from these centroids. 

We suppose a general distance function $f_{\tiny{dis}}$, which is defined to measure the distance between a user to an item, which aims to compute the group affinity $s_{ij}$ of user $i$ and item $j$. The distance metric can be constructed with side information, like users’ social relationships \cite{wang2016social,zhao2017collaborative} or using metric learning techniques \cite{xing2002distance}. However, most datasets do not include such data.  So in this paper we explore a general method to estimate the group affinity. Followed by the idea \cite{lee2013local} of estimating distances with cosine similarities of latent representations, we define the distance of each element $\eqbf{h}_{i} \in $$\eqbf{H}$ and $\mathcal{K}_{k} \in $$\mathcal{K}$, and the distance of each element $\eqbf{g}_{j} \in $$\eqbf{G}$ and $\mathcal{K}_{k} \in $$\mathcal{K}$, respectively as follows:
\begin{equation} \label{gl}
\begin{aligned}
&p_{ik} = f_{\tiny{dis}}(i,k)=\text{cosine}(\eqbf{h}_{i}, \mathcal{K}_{k}) ,  \\
&p_{jk} = f_{\tiny{dis}}(j,k)=\text{cosine}(\eqbf{g}_{j},\mathcal{K}_k)),
\end{aligned}
\end{equation}
where $k \in {1, \cdots, \kappa}$. $p_{ik}, q_{jk}$ characterize distances of latent vectors away from the centroid $\mathcal{K}_k$. Intuitively, users/items in the same group has small discrepancy of distances, we define the group affinity (group similarity) with these distance discrepancies, which composes of $\kappa$ components as follows:
\begin{equation}\label{sim}
s_{ij}^{k}=\sigma(1-\Delta_{ij}^{k}),
\end{equation}
where \begin{equation}
\Delta_{ij}^{k}=|p_{ik}-q_{jk}| 
\end{equation}
for $  k \in \{1, \cdots, \kappa\}$ denotes the distance discrepancy of user $i$ and item $j$ away from the centroid $\mathcal{K}_{k}$, and $\sigma$ is the sigmoid(logistic) function. We can conclude the group affinity is monotone decreasing with the distance discrepancy from Equation \eqref{sim}, that is reasonable because users/items from the same group having smaller distance discrepancies are more affinitive than that from different groups. Furtherly, we define the group affinity $s_{ij}$ as
\begin{equation}\label{sim1}
s_{ij}=\max_{k}s_{ij}^{k}.
\end{equation}
From Equation \eqref{sim1}, we regard the maximum group affinity from $\kappa$ group affinity components as the group affinity of users and items, which is interpretable from the perspective of clustering, which vectors are classified into the nearest class. So we can classify users and items into the same group when the maximum affinity greater than a threshold, thus the maximum affinity with $\kappa$ centroids can be looked as the group affinity.
%The distance between two items can be computed in the same way.

%The set of centroids in $\mathcal{K}$ can also be regarded as a basis of a subspace in the $r-$dimensional space, so it's reasonable to estimate the user-item group affinity $g_{ij}$ (items' abstract types or users' preference types) in the same subspace as follows:
%\begin{equation}
%	g_{ij}=\eqbf{p}_{i}^{T}\eqbf{q}_{j}.
%\end{equation}
Next, we focus on learning hash codes with the obtained group affinity. We assume users and items as binary codes $\eqbf{B}\in \{\pm\}^{r\times n}$ and $\eqbf{D}\in \{\pm\}^{r\times m}$, respectively, then the preference of user $i$ over item $j$ is formulated as follows,
\begin{equation}\label{preference}
\hat{r}_{ij}=s_{ij}\text{sim}_{\tiny{H}}(\eqbf{b}_{i}, \eqbf{d}_{j}),
\end{equation}
where $\text{sim}_{\tiny{H}}(\eqbf{b}_{i}, \eqbf{d}_{j})$ denotes the similarity function of hash codes $\eqbf{b}_{i}$ and $\eqbf{d}_{j}$ defined by their Hamming distances in $r-$dimensional Hamming space, and the similarity function is defined as follows,
\begin{equation}\label{hamming}
\text{sim}_{\tiny{H}}(\eqbf{b}_{i}, \eqbf{d}_{j})=\frac{1}{2}+\frac{1}{2r}\eqbf{b}_{i}^{T}{{\eqbf{d}}_{j}}.
\end{equation}
By substituting Equation \eqref{hamming} into the preference prediction model \eqref{preference}, we can predict ratings with group infinities and similarities of hash codes. We formate the preference as the product of the group affinity and the similarity of hash codes instead of the addition operation, because we tend to learn hash codes in consistent with the group affinity to improve the accuracy for sparse recommendation. Then we derive the following objective of the collaborative group-aware hashing (CGAH) for collaborative filtering,
\begin{equation}
\arg\min_{\eqbf{B}, \eqbf{D}}\mathcal{L}_{\tiny{CGAH}-{CF}}(\eqbf{R}, \eqbf{S}, \eqbf{\hat{R}})+\lambda\Gamma(\eqbf{B}, \eqbf{D}),
\end{equation}
where $\eqbf{\hat{R}}$ is the matrix consisting of the predicted preference $\hat{r}_{ij}$, and the first term is the loss function between ground truth ratings and predicted preference scores, and the second term $\Gamma(\eqbf{B}, \eqbf{D})$ denotes the regularizers (constraints) imposed on hash codes. In this paper, we use the square loss to learn the binary codes in line with the inherent group affinities from ratings. To obtain more compact yet informative hash codes, we impose balance constrains and uncorrelated limitations on hash codes like DCF \cite{zhang2016collaborative}. By substituting the Equation (\ref{hamming}) into Equation (\ref{preference}), the proposed CGAH for collaborative filtering is formulated as follows,
\begin{equation}\label{CGAH}
\begin{aligned}
\arg \min_{\eqbf{B}, \eqbf{D}}&\mathcal{L}_{\tiny{CGAH}-{CF}}(\eqbf{R}, \eqbf{S}, \eqbf{\hat{R}}) =\sum\limits_{{(i,j)}\in\Omega}(r_{ij}-\frac{1}{2}s_{ij}-\frac{1}{2r}s_{ij}\eqbf{b}_{i}^{T}{{\eqbf{d}}_{j}})^{2} \\
s.t. &  \eqbf{B}\in {{\left\{ \pm 1 \right\}}^{r\times n}},\eqbf{D}\in {{\left\{ \pm 1 \right\}}^{r\times m}} \\
&\eqbf{B} \eqbf{1}_n=\eqbf{0},\eqbf{D}\eqbf{1}_m=\eqbf{0}, \eqbf{B}\eqbf{B}^{T}\eqbf{=}n\eqbf{I}_r,\eqbf{D}\eqbf{D}^{T}\eqbf{=}m\eqbf{I}_r,
\end{aligned}
\end{equation}
where  $\Omega=\{(i,j)|r_{ij}\in \eqbf{R}\}$ presents the index of observed ratings in $\eqbf{R}$. 

The idea of CGAH is similar to the compositional coding method CCCF \cite{liu2019compositional}, that denotes hash codes as a weight aggregation of compositional codes. The difference lies in the construction of hash codes. CCCF formulates each hash code as a combination of $\kappa$ hash codes, and the aggregation parameter can be regarded as the group infinity in this work; while CGAH denotes hash codes as binary vectors directly, and the group infinity is joined to learn these hash codes. So hash codes learned from CGAH captures the group information inherited amongst users and items. Besides, the CGAH attains more compact hash codes than CCCF. Because CCCF has to learn $\kappa$ $r$-dimensional hash codes while CGAH only need to learn on $r-$dimensional hash code for each user or item.
%%\vspace{-6pt}
\subsection{CGAH for Content-aware Recommendations} \label{sec33}
\vspace{0.5em}
To alleviate sparse issues in recommender systems, we design CGAH for recommendations by incorporating with sice information, such as users'/items' contexts, images, social links, etc. In this paper, we formulate CGAH with context information of users and items. From Figure \ref{fig0}, we first encode users'/items' content into embeddings $\eqbf{\xi}_{i}$ and $\eqbf{\zeta}_{j}$ of user $i$ and item $j$, and then we concatenate latent vectors from ratings($\eqbf{h}_{i}$ and $\eqbf{g}_{j}$), and embeddings ($\eqbf{\xi}_{i}$ and $\eqbf{\zeta}_{j}$) from contents to calculate their group affinities $s_{ij}$, and finally we learn hash codes by solving the discrete objective of CGAH.

In this work, we encode items'/users' content with denoising auto-encoders (DAE), which can extract robust representations by corrupting the input data by randomly setting some of the input values to zero. The percentage of input nodes which are being set to zero depends on the dimension and the number of input data. Commonly the hidden layer in the middle of DAE is the latent vector we need, and the input layer is the corrupted version of the clean input data. Let the content of user $i$ be denoted as $\eqbf{C}_{i}$, and the content of item $j$ be denoted as $\eqbf{C}_{j}$, and latent vector is the output of the middle layer of DAE denoted as follows,
\begin{equation} \label{dae}
\eqbf{\xi}_{i}=f_{U}(\eqbf{C}_{i}, \Theta_I); \eqbf{\zeta}_{j}=f_{I}(\eqbf{C}_{j}, \Theta_I),
\end{equation}
where $\Theta_{I} =\{\eqbf{W}_{U},\eqbf{b}_{U}\}$ and $\Theta_{J} =\{\eqbf{W}_{I},\eqbf{b}_{I}\}$  are the parameters of encoding networks of two DAE models for users and items, respectively. Actually, we first learn the whole DAE by minimizing the reconstruction error of users' and items' embedding, respectively, and then we fix $\Theta_{I}$ and $\Theta_{J} $ to extract latent vectors $\eqbf{\xi}_{i}$ and $\eqbf{\zeta}_{j}$ by the forward steps.

To calculate the group infinity in latent space, we concatenate latent vectors from MF obtained by Equation \eqref{mf} and DAE extracted by Equation \eqref{dae} together. The concatenation latent vectors are as follows:
\begin{equation}
\eqbf{u}_{i}=[\eqbf{h}_{i}; \eqbf{\xi}_{i}]; \eqbf{v}_{j}=[\eqbf{g}_{j}; \eqbf{\zeta}_{j}].
\end{equation}
Then we take similar steps of Section \ref{sec32} to calculate group affinities $s_{ij}$ of users and items by replacing latent vectors$\eqbf{h}_{i}$ and $\eqbf{g}_{j}$ with $\eqbf{u}_{i}$ and $\eqbf{v}_{j}$, respectively. 

Specifically, we firstly calculate the centroids $\mathcal{K}$ of latent vectors $\eqbf{u}_{i}\in \eqbf{U}$ and $\eqbf{v}_{j}\in \eqbf{V}$ by solving the problem \eqref{group}; then we calculate the distance of each element in $\eqbf{U}$ and each element in $\mathcal{K}$, and the distance of each element in $\eqbf{V}$ each element in $\mathcal{K}$ by the Equation \eqref{gl}, respectively; followed by estimating the group affinity $s_{ij}$ by the Equation \eqref{sim}. 

To solve sparse and cold-start problems, we need to learn hash codes carrying information from both ratings and contents. Motivated by DDL \cite{zhang2018discrete1} which learns hash codes from content data by adding a supervised layer on the output layer of the encoder, and finetunes parameters of the encoder in the learning procedure of hash codes. Thus the objectives of supervised learning for users and items are denoted as follows:
\begin{equation}\label{content}
\begin{aligned}
&\underset{\ \Theta_{I} }{\mathop{\arg \min }}\,\sum\limits_{i=1}^{n}{{{\left\| {{\eqbf{b}}_{i}}-{{\eqbf{\xi}_{i}}} \right\|}_{F}^{2}}}; \\ 
& \underset{\ \Theta_{J} }{\mathop{\arg \min }}\,\sum\limits_{i=1}^{m}{{{\left\| {{\eqbf{d}}_{j}}-{{\eqbf{\zeta}_{j}}} \right\|}_{F}^{2}}}.
\end{aligned}
\end{equation}
By minimizing the objectives in Equation~\eqref{content}, we obtain effective item hash code from content data. Then we formulate the objective of CGAH with ratings and users' and items' content data as follows,
\begin{equation}
\arg\min_{\eqbf{B}, \eqbf{D}, \Theta_{I}, \Theta_{J}}\mathcal{L}_{\tiny{CGAH}}(\eqbf{R}, \eqbf{S}, \eqbf{\hat{R}}, \eqbf{C}_{i}, \eqbf{C}_{j})+\lambda\Gamma(\eqbf{B}, \eqbf{D}, \Theta_{I}, \Theta_{J}),
\end{equation}
where the first term consists of the rating loss function and the two content-based loss functions in Equation \eqref{content}, and it is denoted as follows,
\begin{equation}\label{caghloss}
\mathcal{L}_{\tiny{CGAH}}(\eqbf{R}, \eqbf{S}, \eqbf{\hat{R}}, \eqbf{C}_{i}, \eqbf{C}_{j}) = \mathcal{L}_{\tiny{CGAH}-{CF}}(\eqbf{R}, \eqbf{S}, \eqbf{\hat{R}}) + \lambda_{1}{{{\left\| {{\eqbf{b}}_{i}}-{{\eqbf{\xi}_{i}}} \right\|}_{F}^{2}}} + 
\lambda_{2}{{{\left\| {{\eqbf{d}}_{j}}-{{\eqbf{\zeta}_{j}}} \right\|}_{F}^{2}}},
\end{equation}
where $\lambda_{1}$ and $\lambda_{2}$ are the hyperparameters that weight the importance of three objectives. Similarly, we impose balance constraints and uncorrelated limits on hash codes and obtain the following CGAH objective,
\begin{equation}\label{cgahobj}
\begin{aligned}
\arg\min_{\eqbf{B}, \eqbf{D}, \Theta_{I}, \Theta_{J}}&\mathcal{L}_{\tiny{CGAH}}(\eqbf{R}, \eqbf{S}, \eqbf{\hat{R}}, \eqbf{C}_{i}, \eqbf{C}_{j})\\
&s.t. \eqbf{B} \eqbf{1}_n=\eqbf{0},\eqbf{D}\eqbf{1}_m=\eqbf{0}, \eqbf{B}\eqbf{B}^{T}\eqbf{=}n\eqbf{I}_r,\eqbf{D}\eqbf{D}^{T}\eqbf{=}m\eqbf{I}_r.
\end{aligned}
\end{equation}
%%\vspace{-6pt}
\subsection{Optimization} \label{sec34}
\vspace{0.5em}
In this section, we introduce the way of solving the problems \eqref{CGAH} and \eqref{cgahobj}. As the problem \eqref{CGAH} is similar to the problem \eqref{cgahobj}, and the latter is more complex. We can easily derive the optimization steps of solving problem \eqref{CGAH} by discarding the content objective \eqref{content}. Hence, in this section, we introduce the detailed steps of solving the problem \eqref{cgahobj} for simplicity. 

As the problem (\ref{cgahobj}) is a discrete problem with strict conditions, it is intractable to directly solve objective. In previous work, they generally solve such discrete problems by using the Discrete Coordinate Decent (DCD) \cite{Shen_2015_CVPR}, which learns hash codes with a bitwise way. Another method to solve the discrete problems is proposed in DRMF  \cite{zhang2018discrete}. In contrast to discrete coordinate descent, DRMF firstly finds an upper bound of the original discrete problem by adding variational parameters, and then transforms the upper bound into a series of Binary Quadratic Programming (BQP) sub-problems, and finally updates the entire hash code ${{\mathbf{u}}_{i}}$ in each iteration by directly optimizing these BQP problems with a BQP solver \footnote{https://www.cvxpy.org}. Due to the limit computation of current computers, the BQP solver is time consuming, which leads inefficiency to training the model. 

Consequently, in this paper, we still take a bitwise way to solve the discrete problem. Firstly, we adopt a strategy to solve the problem by softening the balance and uncorrelated constrains with two delegate continuous matrices \cite{zhang2016collaborative}. Let two sets $\mathcal{B}=\{\eqbf{X}\in {{\mathbb{R}}^{r\times n}}\left| \eqbf{X}\eqbf{1}_n=\eqbf{0},\ \eqbf{X}{{\eqbf{X}}^{T}}\eqbf{=}n\eqbf{I}_r \right.\}$, $\mathcal{D}=\left\{ \eqbf{Y}\in {{\mathbb{R}}^{r\times m}}\left| \eqbf{Y}\eqbf{1}_m=\eqbf{0},\ \eqbf{Y}{{\eqbf{Y}}^{T}}\eqbf{=}m\eqbf{I}_r \right. \right\}$. Then we derive the following delegate objective of (\ref{cgahobj}),
\begin{equation}
\begin{aligned}\label{dqcf-soft}
\arg\min_{\eqbf{B}, \eqbf{D}, \Theta_{I}, \Theta_{J}}&\mathcal{L}_{\tiny{CGAH}}(\eqbf{R}, \eqbf{S}, \eqbf{\hat{R}}, \eqbf{C}_{i}, \eqbf{C}_{j})-2\alpha \text{tr}({{\eqbf{B}}^{T}}\eqbf{X})
-2\beta \text{tr}({{\eqbf{D}}^{T}}\eqbf{Y}) \\
&s.t.\ \ \eqbf{B}\in {{\left\{ \pm 1 \right\}}^{r\times n}},\eqbf{D}\in {{\left\{ \pm 1 \right\}}^{r\times m}},
\end{aligned}
\end{equation}
where $\text{tr}({{\eqbf{B}}^{T}}\eqbf{X})$ and $\text{tr}({{\eqbf{D}}^{T}}\eqbf{Y}) $ represent the discrepancies between the binary codes $\eqbf{B}$ and delegated values $\eqbf{X}$, $\eqbf{D}$ and $\eqbf{Y}$, and $\alpha$ and $\beta$ are tuning parameters so that the second and last terms in Eq.\eqref{dqcf-soft} allow certain discrepancy hash codes and delegated vectors. In fact, the two discrepancies are equivalently transformed from $||\eqbf{B}-\eqbf{X}||_{F}^{2}$ and $||\eqbf{D}-\eqbf{Y}||_{F}^{2}$, respectively, because $\text{\emph{tr}}(\eqbf{B}\eqbf{B}^T) = \text{\emph{tr}}(\eqbf{X}\eqbf{X}^T)=nr$ and $\text{\emph{tr}}(\eqbf{D}\eqbf{D}^T) = \text{\emph{tr}}(\eqbf{Y}\eqbf{Y}^T)=mr$. It's worth noting that we do not discard the binary constraint on hash codes $\eqbf{B}$ and $\eqbf{D}$ in the objective (\ref{dqcf-soft}). Then we adopt the following alternating optimization to solve the problem.

\subsection*{\textbf{a. Update $\eqbf{B}$ by fixing $\eqbf{D},\ \eqbf{X},\ \Theta_{I},\ \Theta_{J}$ and $\eqbf{Y}$}} 
\vspace{0.5em}
Since the objective function in the problem~\eqref{dqcf-soft} sums over users independently, we update $\eqbf{B}$ by updating ${{\eqbf{b}}_{u}}$ in parallel by solving the following problem,
\begin{equation}\label{bsub}
\underset{{{\eqbf{b}}_{i}}\in {{\{\pm 1\}}^{r}}}{\mathop{\arg\min }}\,\frac{1}{4r^{2}}\sum\limits_{j\in {{\Omega}_{i}}}s_{ij}^{2}{{{(\eqbf{d}_{j}^{T}{{\eqbf{b}}_{i}})}^{2}}}-\frac{1}{r}\sum\limits_{j\in {{\Omega}_{i}}}{{{g}_{ij}}(r_{ij}-\frac{1}{2}g_{ij})\eqbf{d}_{j}^{T}{{\eqbf{b}}_{i}}}
-2\alpha \eqbf{x}_{i}^{T}{{\eqbf{b}}_{i}} -2\lambda_{1} \eqbf{\xi}_{i}^{T}{{\eqbf{b}}_{i}},
\end{equation}
where ${{\Omega}_{i}}$ is the items set rated by the user $i$. This discrete optimization problem is NP-hard, and thus we adopt the Discrete Coordinate Descent (DCD)~\cite{Shen_2015_CVPR} to update ${{\eqbf{b}}_{i}}$ with a bitwise way. Particularly, let ${{b}_{ik}}$ be the $k$-th bit of ${{\eqbf{b}}_{i}}$, and ${{\eqbf{b}}_{i\bar{k}}}$ be the rest bits of ${{\eqbf{b}}_{i}}$, i.e, ${{\eqbf{b}}_{i}}={{[\eqbf{b}_{i\bar{k}}^{T},{{b}_{ik}}]}^{T}}$, then DCD can update ${{b}_{ik}}$ given ${{\eqbf{b}}_{i\bar{k}}}$. Discarding the terms independent of ${{b}_{ik}}$, the objective in Eq.(\ref{bsub}) is rewritten as
\begin{equation}\label{bsub1}
\arg\underset{{{b}_{ik}}\in \{\pm 1\}}{\mathop{ \min }}\,{{\hat{b}}_{ik}}{{b}_{ik}},
\end{equation}
where ${{\hat{b}}_{ik}}=\frac{1}{4r^{2}}\sum\limits_{j\in {{\Omega}_{i}}}(s_{ij}^{2}{\eqbf{d}_{j\bar{k}}^{T}\ }{{\eqbf{b}}_{i\bar{k}}}-2r{{g}_{ij}}r_{ij}-rg_{ij}){{d}_{jk}}+\alpha {{x}_{ik}}+\lambda_{i} {{\xi}_{ik}}$. The above subproblem can achieve the minimal only if ${{b}_{ik}}$ had the opposite sign of ${{\hat{b}}_{ik}}$. However, if ${{\hat{b}}_{ik}}$ was zero, ${{b}_{ik}}$ would not be updated, so the update rule of ${{b}_{ik}}$ is as follows,
\begin{equation}\label{upb}
{{b}_{ik}}=\text{sgn} \left( \chi\left( -{{{\hat{b}}}_{ik}},{{b}_{ik}} \right) \right),
\end{equation}
where $\chi(x,y)=x$  if $x\ne 0$; otherwise, $\chi(x,y)=y$.

\subsection*{b. Update $\eqbf{D}$ by fixing $\eqbf{B},\ \eqbf{X}, \ \Theta_{I},\ \Theta_{J}$ and $\eqbf{Y}$}
\vspace{0.5em}
Similarly, discarding terms irrelevant to ${{\eqbf{d}}_{j}}$ in Eq~(\ref{dqcf-soft}), we obtain the following subproblem:
\begin{equation}\label{dsub}
\underset{{{\eqbf{d}}_{j}}\in {{\{\pm 1\}}^{r}}}{\mathop{\arg \min }}\,\frac{1}{4r^{2}}\sum\limits_{i\in {{\Omega}_{j}}}s_{ij}{{{(\eqbf{b}_{i}^{T}{{\eqbf{d}}_{j}})}^{2}}}-\frac{1}{2r}\sum\limits_{i\in {{\Omega}_{j}}}{{{g}_{ij}}(2r_{ij}-g_{ij})\eqbf{b}_{i}^{T}{{\eqbf{d}}_{j}}} 
-2\beta \eqbf{y}_{j}^{T}{{\eqbf{d}}_{j}}-2\lambda_{2} \eqbf{\zeta}_{j}^{T}{{\eqbf{d}}_{j}},
\end{equation}
where ${{\Omega}_{j}}$ is the users set that rated item $j$. Similarly, we obtain the following objective with the DCD method,
\begin{equation}\label{dsub1}
\underset{{{b}_{ik}}\in \{\pm 1\}}{\mathop{\arg \min }}\,{{\hat{d}}_{jk}}{{d}_{jk}},
\end{equation}
where ${{\hat{d}}_{jk}}=\frac{1}{4r^{2}}\sum\limits_{i\in {{\Omega}_{j}}}(s_{ij}^{2}\eqbf{b}_{i\bar{k}}^{T}{{\eqbf{d}}_{j\bar{k}}}-2r{{g}_{ij}}r_{ij}-rg_{ij}){{b}_{ik}}+\beta {{y}_{jk}}+\lambda_{2} {{\zeta}_{jk}}$. Similarly, the update rule of ${{d}_{jk}}$ is:
\begin{equation}\label{upd}
{{d}_{jk}}=sgn \left( \chi\left( -{{{\hat{d}}}_{jk}},{{d}_{jk}} \right) \right).
\end{equation}

\begin{algorithm}[!htbp]
	\caption{CGAH}
	\label{alg:dqcf}
	\textbf{Input}: Rating matrix $\eqbf{R}$, Users' and Items' content $\eqbf{T}_{U}$, $\eqbf{T}_{I}$ \\
	\textbf{Parameters}: The number of groups $\kappa$, the length of each hash code $r$, hyper-parameters $\lambda_{1}, \lambda_{2}$, $\alpha$ and $\beta$\\
	\textbf{Output}: Hash codes $\eqbf{B}$ and $\eqbf{D}$
	\begin{algorithmic} [1]%[1] enables line numbers
		\STATE Get group indicators $\eqbf{P}$ and $\eqbf{Q}$ by Eq. \eqref{gl}\\
		Get group affinities $s_{ij}$ of user $i$ and item $j$ by Eq. \eqref{sim1}
		\WHILE{not converged}
		\FOR{$i=1 \cdots n$}
		\WHILE {not converged}
		\FOR{$k=1\cdots r$}
		\STATE Update $b_{ik}$ according to Eq.\eqref{upb} 
		\ENDFOR
		\ENDWHILE
		\ENDFOR
		\FOR{$j=1 \cdots m$}
		\WHILE {not converged}
		\FOR{$k=1\cdots r$}
		\STATE Update $d_{jk}$ according to Eq.\eqref{upd} 
		\ENDFOR
		\ENDWHILE
		\ENDFOR
		\STATE Update $\eqbf{X}$ according to Eq.\eqref{xsub1}
		\STATE Update $\eqbf{Y}$ according to Eq.\eqref{ysub1}
		\STATE Update $\Theta_{I}$ and $\Theta_{J}$ by learning problems in Eq. \eqref{eq:Tsub}
		\ENDWHILE
		\STATE \textbf{return} $\eqbf{B}$, $\eqbf{D}$
	\end{algorithmic}
\end{algorithm}

\subsection*{\textbf{c.}  Update $\eqbf{X}$ by fixing $\eqbf{B}$, $\eqbf{D}, , \ \Theta_{I},\ \Theta_{J}$ and $\eqbf{Y}$}
\vspace{0.5em}
Given $\eqbf{B}$, $\eqbf{D}$, $\eqbf{Y}$, the Eq.\eqref{dsub1} is furtherly transformed as
\begin{equation}\label{xsub}
\underset{\eqbf{X}\in {{\mathbb{R}}^{r\times n}}}{\mathop{{\arg \max}}}\,tr({{\eqbf{B}}^{T}}\eqbf{X}),\ s.t.\ \eqbf{X1}_n=\eqbf{0},\eqbf{X}{{\eqbf{X}}^{T}}\eqbf{=}n\eqbf{I}_r.
\end{equation}
It can be solved with the help of SVD according to~\cite{zhang2016discrete}. Specifically, $\eqbf{X}$ is updated by
\begin{equation}\label{xsub1}
\eqbf{X}=\sqrt{n}[{{\eqbf{U}}_{s}}\ {{\widehat{\eqbf{U}}}_{s}}]{{[{{\eqbf{V}}_{s}}\ {{\widehat{\eqbf{V}}}_{s}}]}^{T}},
\end{equation}
where ${{\eqbf{U}}_{s}}$ and ${{\eqbf{V}}_{s}}$ are respectively stacked by the left and right singular vectors of the row-centered matrix $\eqbf{\bar{B}}: {{\bar{b}}_{ij}}={{b}_{ij}}-\frac{1}{n}\sum\limits_{i=1}^{n}{{{b}_{ij}}}$. ${{\widehat{\eqbf{U}}}_{s}}$ is stacked by the left singular vectors and ${{\widehat{\eqbf{V}}}_{s}}$ can be calculated by Gram-Schmidt orthogonalization, and it satisfies ${{[{{\eqbf{V}}_{s}}\ \eqbf{1}]}^{T}}{{\widehat{\eqbf{V}}}_{s}}=\eqbf{0}$.

\subsection*{d. Update $\eqbf{Y}$ by fixing $\eqbf{B}$, $\eqbf{D}, , \ \Theta_{I},\ \Theta_{J}$ and $\eqbf{X}$}
\vspace{0.5em}
Similarly, we derive the following subproblem of the Eq.\eqref{xsub1} 
\begin{equation}\label{ysub}
\underset{\eqbf{Y}\in {{\mathbb{R}}^{r\times m}}}{\mathop{\text{argmax}}}\,tr({{\eqbf{D}}^{T}}\eqbf{Y}),\ s.t.\ \eqbf{Y1_m}=\eqbf{0},\eqbf{Y}{{\eqbf{Y}}^{T}}\eqbf{=}m\eqbf{I}_r,
\end{equation}
then, $\eqbf{Y}$ is updated by
\begin{equation}\label{ysub1}
\eqbf{Y}=\sqrt{m}[{{\eqbf{P}}_{s}}\ {{\widehat{\eqbf{P}}}_{s}}]{{[{{\eqbf{Q}}_{s}}\ {{\widehat{\eqbf{Q}}}_{s}}]}^{T}}.
\end{equation}
Similarly, ${\eqbf{P}}_{s}$, $\eqbf{Q}_{s}$, ${{\widehat{\eqbf{P}}}_{s}}$, and ${{\widehat{\eqbf{Q}}}_{s}}$ is determined by the row-centered matrix of $\eqbf{D}$.

\subsection*{e. Update  $\Theta_{I}$ ($\Theta_{J}$) by fixing $\eqbf{B}$, $\eqbf{D} $ and $\eqbf{X}, \eqbf{Y}$}
\vspace{0.5em}
We finetunes parameters in encoders of users and items, respectively. The supervised objectives are as follows: 
\begin{equation}\label{eq:Tsub}
\begin{aligned}
&\underset{\Theta_{I} }{\mathop{\arg \min }}\,\sum\limits_{i=1}^{n}{{{\left\| {{\eqbf{b}}_{i}}-f_{U}(\eqbf{C}_{i}, \Theta_{I}) \right\|}^{2}}}; \\
&\underset{\Theta_{J} }{\mathop{\arg \min }}\,\sum\limits_{j=1}^{m}{{{\left\| {{\eqbf{d}}_{j}} - f_{I}(\eqbf{C}_{j}, \Theta_{J}) \right\|}^{2}}}.
\end{aligned}
\end{equation}
Problems~\eqref{eq:Tsub} are supervised learning tasks for users and items, respectively. We choose the stochastic gradient descent (SGD) method to fine-tune parameters of the two encoders, where the gradient descent part is implemented by the Back-propagation algorithm. As ${{\eqbf{b}}_{i}}, {\eqbf{d}}_{j}\in {{\{\pm 1\}}^{r}}$, we choose \emph{tanh} function as the output function of encoders since the output is in the same range $[-1,1]$ with hash codes. Besides,  we choose sigmoid functions as activation functions for hidden layers.  We summarize the optimization method as Algorithm \ref{alg:dqcf}. 

\section{Experiments}
In this section, we evaluate the effectiveness of the proposed CGAH by comparing with the stat-of-the-art hashing based recommendations. As our method can be used for both collaborative filtering denoted as CGAH-CF (introduced in Section \ref{sec32}) and content-aware recommendations denoted as CGAH (introduced in Section \ref{sec33}), so we evaluate the performance of CGAH-CF and CGAH against discrete CF-based methods and discrete content-aware recommendations. Specifically, we present experimental results by answering the following questions:
\begin{itemize}
	\vspace{0.5em}
	\item \textbf{RQ1:} How about the efficiency advantages of hashing-based methods over real-valued methods for the online recommendation?
	\item \textbf{RQ2:} Does the proposed CGAH provide more accurate recommendation with content information by comparing with the state-of-the-art discrete content-aware recommendations?
	\item \textbf{RQ3:} Does the proposed CGAH-CF outperforms the state-of-the-art discrete collaborative filtering methods ?
	\item \textbf{RQ4:} How does the group affinity affect the original matrix factorization?
\end{itemize}

%\vspace{-6pt}
\subsection{Experimental Settings}
\vspace{0.5em}
\subsubsection{Datasets}
We evaluate the performance of our method and the competing baselines on three public datasets: MovieLens-1M\footnote{https://grouplens.org/datasets/MovieLens/}, CDs and Books subsets of Amazon datasets\footnote{http://jmcauley.ucsd.edu/data/amazon/}. The MovieLens dataset contains 1,000,209 user-item interactive data, i.e., ratings within the range of 1 to 5 collected from 6040 users to 3706 movie, which reflects users' preferences over movies on MovieLens website\footnote{https://MovieLens.org}. Amazon datasets contain product reviews and metadata from  Amazon\footnote{https://www.amazon.com/}, which covers 142.8 million reviews (including ratings, text, helpfulness votes) spanning May 1996 - July 2014. In this paer, we use small '5-core' subsets of Amazon datasets for experiments, where each user has rated at least 5 items, and each item has been rated by at least 5 users.

As this paper propose two models for alleviating sparse issue under two settings: CGAH-CF with only ratings available and  CGAH with both ratings and reviews available. So, in this paper, we use the rating datasets of all three datasets to evaluate the effectiveness of CGAH-CF, and use both ratings and reviews on Books and CDs to evaluate the effectiveness of CGAH for content-aware recommendations. For all datasets, we first remove users or items with less than 10 ratings. The statistics of datasets used in this paper are listed in Table \ref{tab1}.

To evaluate the performance of the proposed method on different sparse settings,we randomly choose different proportions' (10\%, 50\%, 90\%) ratings of each user to construct the training datasets, and the remaining ratings to build testing sets. The random selection is carried out 5 times independently, and we report the experimental results as the average values. We evaluate the accuracy performance under NDCG@$k$ \cite{jarvelin2017ir}, which has been widely used for evaluating ranking tasks, especially for recommender systems.
\begin{table}%[!t]
	%\extrarowheight=3pt
	\caption{Statistics of datasets.}
	\small
	\centering
	%\begin{tabular*}{0.41\paperwidth}{@{\extracolsep{\fill}}ccccccc}
	\begin{tabular}{c|c|c|c|c}
		\hline
		%  \hline
		Dataset & \#Users & \#Items& \#Ratings & Sparsity(\%)\\
		\hline
		MovieLens& 6,040& 3,706 & 1,000,209   & 99.97\%\\
		\hline
		CDs& 25,400 & 24,904 & 43,193 & 99.98\%  \\
		\hline
		Books &  603,374&348,957 & 8,575,903 & 99.99\%  \\
		\hline
	\end{tabular}
	\label{tab1}
%	\vspace{-1.5em}
\end{table}

\subsubsection{Evaluation Metric}
As introduced in Section~\ref{sec30}, the task of this paper is to recommend top-$k$ items that users may be interested in. So we choose one of the most common used evaluation metric Normalized Discounted Cumulative Gain(NDCG) \cite{jarvelin2017ir}, to evaluate the performance of predicted ranking lists of items for users. To evaluate the qualities of top-$k$ recommendation results, we need to compute the value of NDCG@$k$, which is widely adopted by many previous ranking-based recommender systems~\cite{koren2008factorization}. 

NDCG@$k$ is a good measure to evaluate the quality of a ranking task, because it takes position significance into account. It is defined as the ratio of Discounted Cumulative Gain(DCG) of a recommended ranked list to DCG of the ideal ranked list,
\begin{equation}
\text{NDCG@}k = \frac{\text{DCG}@k}{\text{iDCG}@k},
\end{equation}
and the ratio takes values in the range of $[0, 1]$. It takes the value `1' when the recommended items' top-$k$ list is the same with the true top-$k$ list. At this moment, the recommendation performance is considered as ideally catering the user's interest. When it takes the value `0', the recommended items' top-$k$ list is totally different from the true top-$k$ list. 

The DCG is defined as the relevance score by dividing it with the log of the corresponding position in the top-$k$ list,
\begin{equation}
\text{DCG@}k = \sum_{i=1}^{k} \frac{2^{\hat{r}_{i}}-1}{\log_{2}(i+1)},
\end{equation}
where the relevance score $\hat{r}_{i}$ is binary, i.e., it takes either `0' or `1'. In predicted recommended top-$k$ list, if a user rated to an item, the corresponding relevance score $\hat{r}_{i}$ takes `1', `0' otherwise. 

iDCG is defined as follows,
\begin{equation}
\text{iDCG@}k = \sum_{i=1}^{k} \frac{2^{r_{i}}-1}{\log_{2}(i+1)},
\end{equation}
where the relevance score $r_{i}$ is also binary. Different from DCG@$k$,  iDCG@$k$ is calculated in the true dataset. In true recommendation top-$k$ list, if a user rated to an item, the corresponding relevance score $r_{i}$ takes `1', `0' otherwise. 
\subsubsection{Baselines}
\vspace{0.5em}
As introduced in Section~\ref{sec2}, we choose two types of comparison methods including the state-of-the-art real-valued (continuous) method MF, three latest competitive hashing-based recommender systems BCCF, DCF, and CCCF for discrete collaborative filtering, and  two kinds of discrete content-aware recommendations DDL and NeuHash-CF. 
\begin{itemize}
	\item \textbf{MF:} Matrix factorization \cite{koren2009matrix} is the most favorite state-of-the-art technique in recommender system, and so there are a lot of variants methods based on MF were put forward in academia and industrial for improving existing recommendations. MF maps users and items in real latent space and then conducts recommendation with continuous representations, and so the online recommendation is not efficient .
	\item \textbf{BCCF:} Binary codes for collaborative filtering \cite{zhou2012learning} is a two-stage hashing-based recommendation, which learn the real latent vectors of users and items, and then quantize them into binary codes. BCCF improves the efficiency of the online recommendation with binary codes, but it suffers poor accuracy performance due to the large information loss in the quantization stage.
	\item \textbf{DCF:} Discrete collaborative filtering \cite{zhang2016discrete} is the state-of-the-art hashing-based recommender systems. Different from BCCF, the DCF directly learn hash codes by optimizing a discrete optimization problem, and then maps the users and items into $r-$dimensional Hamming space. DCF can provide efficient online recommendation, besides, the accuracy is also competitive with continuous methods. However, the DCF performs inferior under sparse scenarios.
	\item \textbf{CCCF:} Compositional coding for collaborative filtering \cite{liu2019compositional} is also a hashing-based recommender system with directly optimizing a discrete problem. CCCF constructs a compositional hashing method with $G$ components for each user and item, respectively.		
	\item \textbf{DDL:} Discrete deep learning \cite{zhang2018discrete1} is a content-aware hashing method for solving cold start and spare problem. DDL also collaborates interactive data and content data to learn hash codes of users and items. The differences between DDL and CGAH include the group infinity proposed in CGAH and the capability of solving the user cold start problem.
	\item \textbf{NeuHash-CF:} 	Content-aware Neural Hashing \cite{hansen2020content} was formulated as an autoencoder architecture consisting of two joint hashing components for generating user and item hash codes.
\end{itemize}
\subsubsection{Hyper-parameter Settings}
We tune the optimal hyper-parameters of our algorithm by the grid search method. Specifically, for the proposed CGAH, we search the optimal hyper-parameters $\lambda$, $\lambda_{1}$ and $\lambda_{2}$ over $\{1e^{-4}, 1e^{-3}, 1e^{-2}, 1e^{-1}, 1e^{1}, 1e^{2}\}$] on each dataset. The dimension of discrete/continuous representations is set to $r = 20$ by default. The number of groups for users' preference over items is set to $\kappa = 10$. Finally, we set $\lambda=0.1$ for CGAH-CF, and set $\lambda_{1}=0.2$ and $\lambda_{2}=0.1$ for CGAH for content-aware recommendations. For other baselines, we use the default optimal hyper-parameters in their public codes to evaluate their performances on each dataset.
%%\vspace{-6pt}
\subsection{Experimental Results and Analysis}
\vspace{0.5em}
%%\vspace{-6pt}
\subsubsection{Efficiency comparison with baselines (\textbf{RQ1}) }

%As discussed in other hashing-based methods \cite{zhang2016discrete,zhang2017discrete,liu2019compositional}, CGAH maps users and items in $r-$dimensional Hamming space, and so the online recommendation can be significantly accelerated with the fast similarity search by XOR bit operations. Table \ref{tab2} shows the efficiency performance of the online recommendation on both Amazon dataset  and MovieLens by comparing the continuous method MF. Due to continuous methods obtain real-valued representations, and so continuous methods perform similar for the online recommendation stage, and thus we choose the representative MF for comparison. Similarly, hashing-based methods performs similar in online recommendation as introduced in Section \ref{sec1}. Table \ref{tab2} shows the significant superiority of hashing-based method CGAH for providing efficient online recommendation.

As discussed in other hashing-based methods \cite{zhang2016discrete,zhang2017discrete,liu2019compositional}, recommendations with hash codes can be significantly accelerated with the fast similarity search by XOR bit operations. Due to all continuous methods obtain real-valued vectors, and the online recommendation stage consists of calculating the predicted ratings based on the obtained continuous vectors and ranking these ratings to recommend top-$k$ items for specific users. Hence, the online recommendation stage are the same for all continuous methods, and thus we choose the representative MF for comparison. Table \ref{tab2} shows the efficiency comparisons of the online recommendation on Books dataset by comparing hashing-based methods with the continuous method MF. 

From Table \ref{tab2}, we find that the proposed CGAH provide the fastest online recommendation among all hashing-based methods, and it takes least storage (the least compress ratio). Compared with CCCF, our method attains one $r-$dimension hash codes for each user/item; while CCCF obtain $\kappa$ hash codes of $r-$dimension for each user/item. That's why CCCF costs most time and storage for online recommendation. For BCCF, it learns hash codes without balance constraint. If we save hash codes as sparse format, it will cost more than balance hash codes obtained from DCF and our method. Compared with DCF, our method gets similar hash codes for similar users/items (with greater group affinity), while DCF may get different hash codes for similar users/items. That's the reason that the proposed CGAH is efficient for online recommendations.

\begin{table}[!htbp]
	%\extrarowheight=3pt
	\caption{Speedup ratios(\%) of hashing-based frameworks to the continuous MF on Books dataset, and the storage compress ratios(\%) of items' representations}
	%		\vspace{-6pt}
	%	\small
	\centering
	%\begin{tabular*}{0.41\paperwidth}{@{\extracolsep{\fill}}ccccccc}
	\centering
	\begin{tabular}{c|ccc|c}
		\toprule
		\multirow{2}{*}{} & \multicolumn{3}{c|}{Speedup ratios(k=10, 50, 100)} &\multirow{2}{*}{Compress ratios} \\ & 10 & 50&100&  \\ \hline
		CCCF& 35.50 & 29.60& 21.69 & 30.21 \\ \hline
		BCCF & 17.65 & 14.38& 10.07& 15.53 \\ \hline
		DCF& 14.17 & 10.31& 9.25 & 10.29  \\ \hline
		\textbf{CGAH}& \textbf{12.82} & \textbf{10.32}&\textbf{8.05}&\textbf{ 8.90}  \\ \hline
		\toprule
	\end{tabular}
	\label{tab2}
%	\vspace{-1.5em}
\end{table}
%\vspace{-6pt}

%%\vspace{-6pt}
\subsubsection{The Effectiveness of CGAH for Content-aware Recommendations(\textbf{RQ2})}
\vspace{0.5em}
As the objective is to solve the sparse issue in recommendations, thus, in this part we randomly choose 10\% ratings of each user as training dataset, and the remaining 90\% ratings of each user as testing set. Figure \ref{fig3} shows the experimental results of CGAH for providing top-$k$ recommendations with content information by comparing with other two content-aware hashing-based recommendations NeuHash-CF and DDL on CDs and Books, respectively. 

From the Figure \ref{fig3}, we find our method outperforms other two competing baselines in the sparse setting, and we have the following findings:
\begin{figure}[!htbp]
	\centering
	\includegraphics[width=1\linewidth]{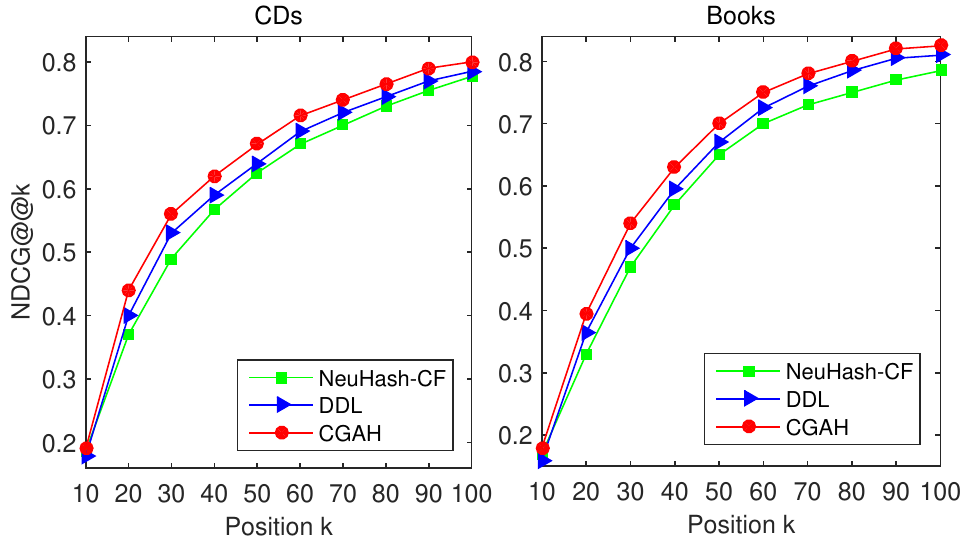}
	%	\vspace{-2em}
	\caption{The performance comparison of discrete content-aware recommendations on CDs and Books with the sparse setting 10\%}\label{fig3}
	%		\vspace{-6pt}
\end{figure}
\begin{itemize}
	\item \textit{The group information carried by hash codes can improve the accuracy.} We know that DDL and NeuHash-CF are hashing methods collaborated with content information. DDL and NeuHash-CF are designed to learn representations from ratings and content, which aims to learn models to fit ratings; while CGAH is formulated to adapt both the inherent group information (from content and ratings) and the observed ratings. So the CGAH attain binary codes with group information, and thus it can provide more accurate recommendations under the very sparse setting 10\%. 
	\item \textit{The group information helps to enhance the representation ability of hash codes.} Despite that hashing-based recommender systems can speed up the online recommendation by the fast similarity search in Hamming space, but each bit naturally carries much less information than each continuous dimension, which leads to poor recommendation accuracy. By extracting group information hided in ratings, CGAH obtains hash codes in consistent with the group affinity of users and items. That's why CGAH can achieve the best performance.
\end{itemize}

\subsubsection{The Effectiveness of CGAH for Collaborative Filtering (CGAH-CF) (\textbf{RQ3}) }
\vspace{0.5em}
%Figure \ref{fig2} shows  the accuracy performance of CGAH-CF by comparing with discrete CF methods on CDs dataset. The experiments evaluate CGAH-CF outperforms other three discrete CF methods: BCCF, DCF and CCCF under different sparse settings 10\%, 50\% and 90\%. Because BCCF, DCF and CCCF are designed to fit the observed ratings while CGAH-CF are modeled to fit both the inherent group information and the observed ratings. In addition, hash codes with CGAH-CF save some group information. That's why CGAH can achieve the best performance under various sparse settings .

\begin{figure}[!htbp]
	\centering
	\includegraphics[width=1\linewidth]{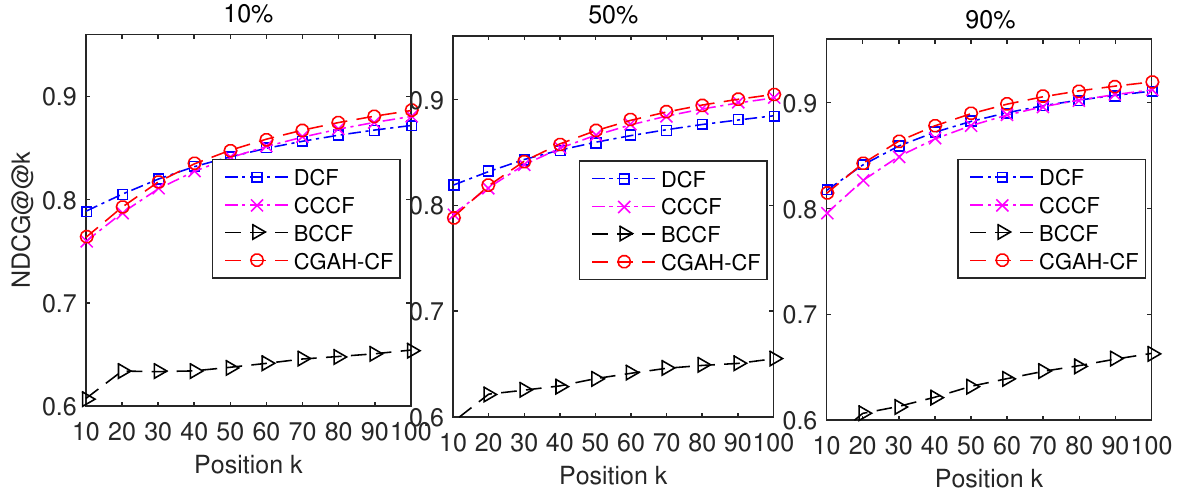}
	%		\subfigure{\includegraphics[width=1\linewidth]{mzf}}
	\caption{The accuracy performance for collaborative filtering on MovieLens with different sparsity levels(10\%, 50\%, 90\%).}\label{fig1}
	%%\vspace{-1em}
\end{figure}

\begin{figure}
	\centering
	\includegraphics[width=1\linewidth]{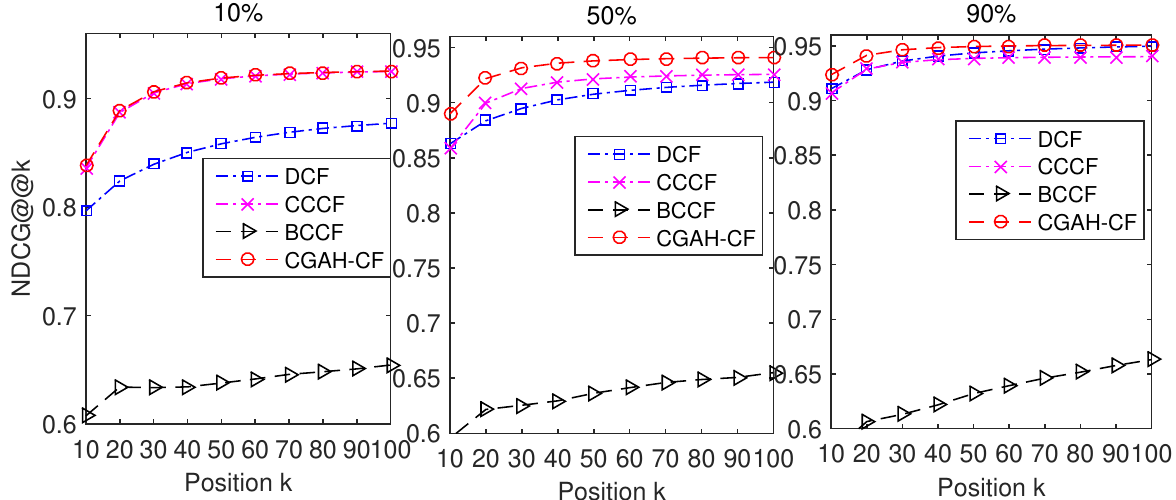}
	\caption{The accuracy performance for collaborative filtering on CDs dataset with different sparse settings (10\%, 50\%, 90\%).}\label{fig2}
	%%\vspace{-1em}
\end{figure}

Figure \ref{fig1} and Figure \ref{fig2} show the accuracy performance by comparing with discrete CF methods on MovieLens and CDs datasets, respectively. The two figures indicate the proposed CGAH-CF outperforms other three discrete CF methods: BCCF, DCF and CCCF under different sparse settings 10\%, 50\% and 90\%, and we have the following findings from the results. 
\begin{itemize}
	\item \textit{The group information carried by hash codes can improve the accuracy.} We know that CGAH-CF, DCF and CCCF are hashing methods by directly optimizing discrete problems. DCF and CCCF are designed to fit the observed ratings; while CGAH-CF is formulated to adapt both the inherent group information and the observed ratings. So the CGAH can attain binary codes with group information, and thus it can provide more accurate recommendation under various sparse settings. 
	\item \textit{The group information helps to enhance the representation ability of hash codes.} As CGAH-CF can obtain hash codes that are consistent with the group affinities. Hence, hash codes should be more effective to represent both user-item interactions and user-item group affinities. However, other discrete CF methods are not capable of representing group information with hash codes. That's why the proposed CGAH-CF performs better than baselines.
\end{itemize}
%	\item \textbf{}
%\end{itemize}

\subsubsection{The Effectiveness of the Group Affinity for Matrix Factorization(\textbf{RQ4}) }
\begin{figure}[!h]
	\centering
	%		\vspace{-1em}
	\includegraphics[width=1\linewidth]{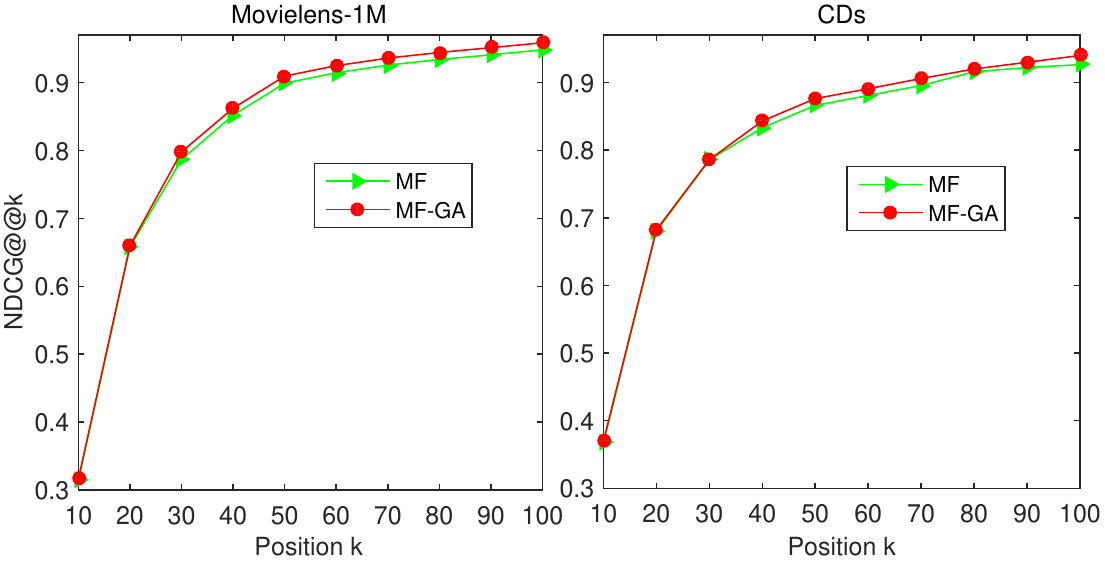}
	\caption{Ablation studies of the group affinity on MF with the sparse setting 90\%.}\label{fig4}
%	\vspace{-1.5em}
\end{figure}

\vspace{0.5em}
To evaluate the effectiveness of group affinities developed in the proposed CGAH, we do ablation studies on the state-of-the-art MF model. MF trained together with group affinities is denoted as MF-GA. The experimental results shown in Figure \ref{fig4} and Table \ref{tab4} tell us the performance of MF-GA is better than the original MF model. For MF-GA, we only add the group affinities calculated by Equation \ref{sim1} on the original MF, and other components in MF and the learning algorithm keep the same with MF. The objective of MF is introduced in the problem \eqref{mf}, and the objective of MF-GA is as follows.
\begin{equation}\label{mf-ga}
\arg \min_{\eqbf{H}, \eqbf{G}} \mathcal{L}_{\tiny{MF-GA}}(\eqbf{R}, \eqbf{S}, \eqbf{H}^{T}\eqbf{G})+\lambda \Gamma(\eqbf{H}, \eqbf{G}),
\end{equation}
where $\eqbf{S}$ is the group affinity computed by Equation \eqref{sim} and \eqref{sim1}. 

 Figure \ref{fig4} shows the performance of MF-GA is better than the original MF. The superiority becomes more evident when two recommender systems recommend more items to users, as the difference of NDCG@$k$ between MF-GA and MF becomes obvious with the rise of $k$. Which indicates that the group affinity is effective to provide more accurate recommendations.

\begin{table}%[!t]
	%\extrarowheight=3pt
	\caption{The effectiveness of group affinity for MF. (MF-GA: MF with group affinity) on Books dataset}
	%		\vspace{-6pt}
%	\vspace{-1em}
	%	\small
	\centering
	%\begin{tabular*}{0.41\paperwidth}{@{\extracolsep{\fill}}ccccccc}
	\begin{tabular}{c|c|c|c|c|c}
		\hline
		%  \hline
		NDCG@k & 10 & 20&30&40&50 \\
		\hline
		MF& 0.3286& 0.6390& 0.7705& 0.8402&0.8754\\
		\hline
		\textbf{MF-GA} &\textbf{0.3428}& \textbf{0.6492}&\textbf{0.7832}& \textbf{0.8528}& \textbf{0.8851}\\
		\hline
		%		Movies &  0.9310& 0.0716&$13$ times $\uparrow$\\
		%		\hline
	\end{tabular}
	\label{tab4}
%	\vspace{-1.5em}
\end{table}
%\vspace{-6pt}
%%\vspace{-6pt}
\section{Conclusion}
\vspace{0.5em}
%This paper proposes the Collaborative Group-Aware Hashing (CGAH), which not only gains the group information from ratings and content data, but also obtains a shared codebook of users and items, which is essential for enhancing the representation capability of hash codes. Specifically, CGAH firstly extracts the group information of users and items in $r-$dimensional latent space, and then represents each user/item with these group centroids. The group centroids serve as a basis of in the $r$-dimensional latent space, and then the group affinity is calculated as a function of the distance discrepancy in the latent space, and finally the preference is modeled as the product of the group infinity and the similarity of hash codes in Hamming space. By learning hash codes with the inherent group information, CGAH can obtain more robust representation than other discrete methods under sparse environments. Extensive experiments on three public datasets show the superiority of CGAH for collaborative filtering and recommendation with side information over the state-of-the-art discrete CF methods under both dense and sparse settings.

This paper proposes a framework named Collaborative Group-Aware Hashing (CGAH) for both collaborative filtering (CGAH-CF) and content-aware recommendations (CGAH), which not only gains the group information from ratings, but also obtains a shared codebook of users and items, which is essential for enhancing the representation capability of hash codes. To the best of our knowledge, this is the first paper to explicitly capture the group affinities and integrate them with ratings into a model, which makes it possible to learn more effective representations of users and items for accurate recommendations.

Specifically, CGAH firstly extracts the group information of users and items in $r-$dimensional latent space, and then represents each user/item with these group centroids. The group centroids serve as a basis in the $r$-dimensional latent space. Then the group affinity is calculated by a function of distance discrepancy in the latent space. Finally, the preference is modeled as the product of the group affinity and the similarity of hash codes in $r-$dimensional Hamming space. By learning hash codes with the inherent group information, CGAH obtains more effective hash codes than other discrete methods under different sparse environments. Extensive experiments on three public datasets show the superiority of CGAH for discrete collaborative filtering and content-aware recommendations over competing baselines.

%\section*{Acknowledgments}

\section*{Acknowledgments}
%	\vspace{0pt}
This work is supported by the Major Project for New Generation of AI under Grant No. 2018AAA0100400, the National Natural Science Foundation of China under Grant No. 62002052 and the Sichuan Science and Technology Program under Grant No. 2022YFG0189.
%\balance
\bibliographystyle{ACM-Reference-Format}
\bibliography{ijcai21}
%\bibliographystyle{ACM-Reference-Format}
%\bibliography{sigir}
%%

\end{document}